\documentstyle[twoside,fleqn,espcrc2,equations,epsfig]{article}

\newcommand{\AmS}{{\protect\the\textfont2
  A\kern-.1667em\lower.5ex\hbox{M}\kern-.125emS}}

\begin{document}
\noindent
\hspace*{11.6cm}
KEK-TH-506\\
\noindent
\hspace*{11.3cm}
SNUTP 97-008\\
\noindent
\hspace*{11.5cm}
YUMS 97-001\\
\noindent
\hspace*{11.4cm}
(January 1997)\\
\vskip1.5cm
\begin{center}
{\bf\Large Measuring of $|V_{ub}|$ in the forthcoming decade\footnote{
Talk is given at `The 4th KEK Topical Conference on Flavor Physics'.
The work  was supported 
in part by the KOSEF, Project No. 951-0207-0F08-2,
in part by the BSRI Program, Project No. BSRI-97-2425,
in part by CTP of SNU, and in part by the COE fellowship of Japanese
Government.}}\\
\vskip0.5cm

{\Large C.S. Kim\footnote{kim@cskim.yonsei.ac.kr, cskim@kekvax.kek.jp}}\\
\vskip0.5cm

{\it Department of Physics, Yonsei University,
	Seoul 120-749, Korea\footnote{
Present address till Feb. 1997: Theory Division, KEK, Tsukuba, 
Ibaraki 305, Japan. }}\\
\vskip0.5cm
(\today)
\vskip1.5cm
	
{\bf abstract}\\	
\vskip0.5cm

\begin{minipage}{13cm}
I first introduce the importance of measuring $V_{ub}$ precisely.
Then, from a theoretician's point of view, 
I review (a) past history, (b) present trials, and (c) possible
future alternatives on measuring $|V_{ub}|$ and/or $|V_{ub}/V_{cb}|$. 
As of my main topic, I introduce a model-independent method, 
which predicts
$\Gamma(B \rightarrow X_u l \nu) / \Gamma(B \rightarrow X_c l \nu)
\equiv (\gamma_u / \gamma_c) \times | V_{ub} / V_{cb}|^2 \simeq (1.83 \pm 0.28) 
\times |V_{ub} / V_{cb}|^2$ and 
$|V_{ub}/V_{cb}| \equiv (\gamma_c / \gamma_u)^{1/2} \times 
[{\cal{B}}(B \rightarrow X_u l \nu)/{\cal{B}}(B \rightarrow X_c l \nu)]^{1/2}
\simeq (0.74 \pm 0.06) \times [{\cal{B}}(B \rightarrow X_u l \nu)/
{\cal{B}}(B \rightarrow X_c l \nu)]^{1/2}$,
based on the heavy quark effective theory. 
I also explore the possible experimental options to separate 
$B \rightarrow X_u l \nu$ from the dominant 
$B \rightarrow X_c l \nu$: the measurement of
inclusive hadronic invariant mass distributions, and 
the `$D - \pi$' (and `$K - \pi$') separation conditions. 
I also clarify the relevant experimental backgrounds.
\end{minipage}
\end{center}
\newpage
\title{Measuring of $|V_{ub}|$ in the forthcoming decade }

\author{C.S. Kim\address{Department of Physics, Yonsei University,
	Seoul 120-749, Korea}
\thanks{Present address till Feb. 1997: Theory Division, KEK, Tsukuba, 
	Ibaraki 305, Japan. 
	The work  was supported 
	in part by the KOSEF, Project No. 951-0207-0F08-2,
	in part by the BSRI Program, Project No. BSRI-97-2425,
	in part by CTP of SNU, and in part by the COE fellowship of Japanese
	Government.}}
       

\begin{abstract}
I first introduce the importance of measuring $V_{ub}$ precisely.
Then, from a theoretician's point of view, 
I review (a) past history, (b) present trials, and (c) possible
future alternatives on measuring $|V_{ub}|$ and/or $|V_{ub}/V_{cb}|$. 
As of my main topic, I introduce a model-independent method, 
which predicts
$\Gamma(B \rightarrow X_u l \nu) / \Gamma(B \rightarrow X_c l \nu)
\equiv (\gamma_u / \gamma_c) \times | V_{ub} / V_{cb}|^2 \simeq (1.83 \pm 0.28) 
\times |V_{ub} / V_{cb}|^2$ and 
$|V_{ub}/V_{cb}| \equiv (\gamma_c / \gamma_u)^{1/2} \times 
[{\cal{B}}(B \rightarrow X_u l \nu)/{\cal{B}}(B \rightarrow X_c l \nu)]^{1/2}
\simeq (0.74 \pm 0.06) \times [{\cal{B}}(B \rightarrow X_u l \nu)/
{\cal{B}}(B \rightarrow X_c l \nu)]^{1/2}$,
based on the heavy quark effective theory. 
I also explore the possible experimental options to separate 
$B \rightarrow X_u l \nu$ from the dominant 
$B \rightarrow X_c l \nu$: the measurement of
inclusive hadronic invariant mass distributions, and 
the `$D - \pi$' (and `$K - \pi$') separation conditions. 
I also clarify the relevant experimental backgrounds.
\end{abstract}

\maketitle

\section{INTRODUCTION}

A precise determination of Cabibbo Kobayashi Maskawa (CKM) matrix
elements \cite{ckm} is the most important goal of the forthcoming
$B$-factories \cite{B-fact}, CLEO-III, KEK-B, SLAC-B, HERA-B, LHC-B.
Their precise values are urgently needed for analyzing CP-violation and for
testing the Standard Model (SM) through the unitarity relations
among them \cite{rosner}. 
Furthermore, the accurate knowledge of these matrix elements can
be useful in relating them to the fermion masses and also in the searches 
for hints of new physics beyond the SM \cite{barbieri}.
Even precision measurements on Top-physics will be affected, because
the value of $V_{td}$ is related to $V_{ub}$ through the unitary relation.

The CKM matrix element $V_{ub}$ is important to the SM description  
of CP-violation. If it were zero, there would be no CP-violation from 
the CKM matrix ({\it i.e.} in the SM), and we have to seek 
for other sources of CP violation in $K_{L} \rightarrow \pi\pi$. 
Observations of semileptonic $b\rightarrow u$ transitions by the CLEO 
\cite{cleo} and ARGUS \cite{argus}  imply that $V_{ub}$ is  indeed nonzero, 
and it is important to extract the modulus $|V_{ub}|$ from semileptonic 
decays of $B$ mesons as accurately as possible.  
  
\section{OVERVIEW OF MEASURING $|V_{ub}|$}

\subsection{Past history}

Historically, the charged lepton energy spectrum ($d \Gamma / d E_l$)
has been measured, and the $b\rightarrow u$ events are selected from the high 
end of the charged lepton energy spectrum.  This method is applied to both  
inclusive and exclusive semileptonic $B$ decays.  
However, this cut on $E_l$ is not very effective, since only less than
$10 \%$ of $b\rightarrow u$ events survive this cut at the $B$ meson 
rest frame. (In the future asymmetric $B$-factories with boosted $B$ mesons, 
much less than $10 \%$ of $b\rightarrow u$ events would survive the $E_l$ 
cut over the $b \rightarrow c$ threshold.)
We also note that the dependences of  the lepton energy spectrum  on 
perturbative and non-perturbative QCD corrections \cite{kuhn,hqet} 
as well as on the unavoidable specific model parameters 
({\it e.g.} the parameter $p_{_F}$ of the ACCMM model \cite{accmm}) are 
strongest at the end-point region, 
which makes the model-independent determination of $|V_{ub}/V_{cb}|$ 
almost impossible from the inclusive distribution of
$d \Gamma / d E_l$.

For exclusive $B \rightarrow X_u l \nu$ decays, the application of 
heavy quark effective theory (HQET) is very much limited, since $u$-quark
is not heavy compared to $\Lambda_{QCD}$. And the theoretical predictions
for the required hadronic matrix elements are  largely
different depending on which model we use, as can be seen in the following,
as an example for $\bar B^0 \rightarrow \rho^+ l^- \bar\nu$,
\begin{eqalignno}
\gamma_{\rho} &\equiv 
{\Gamma_{theory}(\bar B^0 \rightarrow \rho^+ l^- \bar\nu) 
\over |V_{ub}|^2}  \label{eq1} \\
&=~~8.3 \times 10^{12}/sec~,~~~(\cite{isgw}) \nonumber\\
&= 32.9 \times 10^{12}/sec~,~~~(\cite{KS}) \nonumber\\
&= 18.7 \times 10^{12}/sec~.~~~~(\cite{WSB}) \nonumber
\end{eqalignno} 
See also Fig. 1 for the explicit model dependence on the value of
$\gamma_\rho$.
There are certainly many more available models than the listed above.
And every one of them is based on a few reasonable assumptions.
However, such assumptions of one model are in general exclusive 
to the assumptions of the other models, {\it e.g.} monopole dominance
or multipole dominance. And the usual practice of selecting
a few models and averaging the few chosen results is physically
groundless. These model dependences are not like the statistical errors.

\subsection{Present trials}

Measurement of exclusive charmless semileptonic decays can put constraints
on the models and therefore restrict the model dependence in principle, 
if the ratio of rates for $\pi\ell\nu$ and $\rho\ell\nu$ as well
as the $q^2$ dependence of the form-factors are precisely measured. 
CLEO has recently succeeded in measuring the branching
ratio $BR(B \rightarrow \rho\ell\nu)$ \cite{pilnuex}.

A neutrino reconstruction technique is used: The neutrino energy and 
momentum are determined by evaluating the missing momentum and energy in the 
entire event:
\begin{eqalignno}
E_{miss} &= 2E_{beam}-\sum_{i}E_i, \nonumber \\
\overrightarrow{p}\!_{miss} &=  \sum_{i}\overrightarrow{p}\!_i .
\end{eqalignno}
More criteria are imposed to guard against events with
false large missing energies: First, the net charge is required to be zero.
Secondly, events with two identified leptons (implying two neutrinos) are
rejected. Leptons are required to have momenta greater than 1.5 GeV in
the case of $\pi\ell\nu$ and greater than 2.0 GeV in the case of
$\rho\ell\nu$. In addition, the candidate neutrino mass is calculated as
\begin{eqalignno}
M^2_{\nu}=E^2_{miss}-\overrightarrow{p}^2\!\!_{miss}.
\end{eqalignno}
Candidate events containing a neutrino are kept if 
$M^2_{\nu}/2E_{miss}< 300 $ {\rm MeV}. Then the semileptonic $B$ decay
candidates ($\pi^o,~\pi^+,~\rho^o,~\omega^o,~\rho^+)\ell\nu$ are reconstructed
using the neutrino four-vector found from the missing energy 
measurement. 
The beam constrained invariant mass, $M_{cand}$ is defined as
\begin{eqalignno}
M^2_{cand}=E^2_{beam}-\left(\overrightarrow{p}\!_{\nu}+
\overrightarrow{p}\!_{\ell}
+\overrightarrow{p}\!_{(\pi{\rm ~or~}\rho})\right)^2
\end{eqalignno}
and with the use of the neutrino four-vector is essentially the same as any 
other full $B$ reconstruction analysis done at the $\Upsilon(4S)$.

However, it is often difficult to prove that a $\pi\pi$ system indeed 
is dominantly from resonant $\rho$ \cite{moneti}. 
CLEO attempts to show $\rho$ dominance by plotting the
$\pi^+\pi^-$ and $\pi^+\pi^o$ summed mass spectrum. They also
show a test case of $\pi^o\pi^o\ell\nu$, which cannot be $\rho$, since
$\rho^o$ cannot decay to $\pi^o\pi^o$. There is an enhancement in the
$\pi^+\pi^-$ plus $\pi^+\pi^o$ sum, while the $\pi^o\pi^o$ shows a 
relatively flat spectrum that is explained by background. 
CLEO proceeds by assuming they are seeing purely resonant decays in the
vector channel.

Experimental importance here are that CLEO-III can be a powerful 
$B$-factory with
possibly more than $10^5$ fully reconstructed semileptonic decay events. 
However, they have to find 
the way to avoid the most difficult problem; the large model
dependence for exclusive $b \rightarrow u$ decays, as shown in Eq. (1),
and as explained in previous Section 2.1. 
Better options shown in following Sections  should be seriously
pursued by CLEO-III experiment.

\subsection{Future alternatives}

(a) The possibility of measuring $|V_{ub}|$ via
non-leptonic decays of $B$ mesons to exclusive two meson final states
\cite{soni} has been theoretically explored.
To avoid the theoretical difficulties of non-spectator decay
diagrams, only those final states must be chosen in which no
quark and antiquark pair has the same flavor.
Within the factorization approximation and after considering 
the final state interactions, exclusive two body decay modes
of $B$ mesons would certainly be  worth of full investigation.

(b) It has also been  suggested  that the measurements of
hadronic invariant mass spectrum \cite{kim-ko,cskim} as well as 
hadronic energy spectrum \cite{bouzas}
in the inclusive $B \rightarrow X_{c(u)} l \nu$ decays can be
useful in extracting $|V_{ub}|$ with better theoretical understandings.
Experimentally, the hadron energy spectrum in semileptonic $B$ decays
may be measured schematically as follows \cite{rey}:
Working at the $\Upsilon(4S)$ resonance,
which decays into $B \bar B$, one requires one of the $B$-mesons to
decay semileptonically and the other one hadronically. In the case of
a symmetric $B$-factory, like CLEO, the energy of the hadrons stemming
from the semileptonically decaying $B$-meson can be obtained by
measuring the  total  energy of all the hadrons in the final state
and then subtracting $m_{\Upsilon(4S)}/2$.
In case of asymmetric $B$ factories, the hadron
energy spectrum is harder to measure. One way is
to reconstruct in a first step the whole $\Upsilon(4S)$ decay
in its rest frame and then perform the analysis just described
for the symmetric case.
After imposing a relatively high lower--cut at 
$E_{had}=1$ GeV (in order to avoid
the region of phase space where the range in the invariant
hadronic mass is too narrow to invoke quark-hadron duality),
a much larger fraction ($\sim 25\%$) of the  $b \to u \ell^- {\overline \nu}$ 
events is captured in the remaining window $1 \,
\mbox{GeV} \le E_{had} \le m_D$
than in the lepton spectrum endpoint analysis.

(c) The measurement of ratio $|V_{ub}/V_{ts}|$ from the differential
decay widths of the processes $B \rightarrow \rho l \nu$ and
$B \rightarrow K^* l \bar l$ by using $SU(3)$-flavor symmetry and
the heavy quark symmetry has been also proposed \cite{sanda}.
Then the ratio $|V_{ub}|^2/|V_{ts}|^2$ is extracted as
\begin{eqalignno}
\frac{|V_{ub}|^2}{|V_{ts}|^2} &\propto
\frac{{q^2}_{max}^{B \rightarrow K^*}}{{q^2}_{max}^{B \rightarrow \rho}} 
\label{eqn:main} \\ \times
\left[\frac{d \Gamma(B \rightarrow \rho l \nu)}{dq^2}\right]_{q^2_{max}}
 &/
\left[\frac{d \Gamma(B \rightarrow K^* l \bar l)}{dq^2}\right]_{q^2_{max}}.
\nonumber
\end{eqalignno}
In the limit $q^2 \rightarrow q^2_{max}$, the $q^2$
distributions vanish due to the phase space suppression.
In fact, CLEO collaboration has rather accurately
determined the value of $|V_{cb}| \cdot f(q^2_{max})$
for the process $B \rightarrow D^* l \bar{\nu}$ \cite{cleo1}
by extrapolating the $q^2$ distribution.
In the similar manner, the right-hand side of Eq. (\ref{eqn:main})
can be determined by experiments.
There has also been a recent theoretical progress on the exclusive 
$b \rightarrow u$ 
semileptonic decay form factors using the HQET-based scaling laws 
to extrapolate 
the form factors from the semileptonic $D$ meson decays \cite{hqet-based}.

(d) It is urgently important that all the available methods have to be 
thoroughly explored to measure the most important CKM matrix element 
$V_{ub}$ as accurately as possible in the forthcoming $B$-factories. 
In future asymmetric $B$-factories (or in hadronic $B$-factories) 
with microvertex detector, 
the hadronic invariant mass spectrum 
(or `$D - \pi$', `$K - \pi$' separation conditions) offer
alternative ways to select $b \rightarrow u$ transitions that are much more
efficient than selecting the upper end region of the lepton energy spectrum, 
with much less theoretical uncertainties \cite{cskim}.
Then we can use the simple relation, 
\begin{eqalignno}
{|V_{ub}| \over |V_{cb}|} = \left({\gamma_c \over \gamma_u}\right)^{1/2}
\times \left[{{\cal{B}}(B \rightarrow X_u l \nu) \over
{\cal{B}}(B \rightarrow X_c l \nu)}\right]^{1/2} 
\end{eqalignno}
where $\gamma_u$, $\gamma_c$ and $\gamma_u/\gamma_c$ can be calculated 
model-independently within the HQET. 
However, we note that  the individual exclusive decay width, $e.g.$ 
$\gamma_\rho$ or $\gamma_\pi$, cannot be predicted model-independently, 
as shown in Eq. (1) and in Fig. 1.
We will give in more detail on this alternative method 
in the following Section.

\section{MEASURING $|V_{ub}|$ MODEL INDEPENDENTLY}

\subsection{Theoretical proposal}

Over the past few years, a great progress has been achieved 
in our understanding of {\it inclusive} semileptonic decays 
of heavy mesons \cite{hqet}, especially in the lepton energy spectrum.
However, it turns out that the end-point region of the lepton energy 
spectrum cannot be described by 
$1/m_{_Q}$ expansion. Rather, a partial resummation of  $1/m_{_Q}$ 
expansion is required \cite{resum}, closely analogous to the leading
twist contribution in deep inelastic scattering, which could bring about 
significant uncertainties and presumable model dependences.

Even with a theoretical breakdown near around the end-point region of lepton
energy spectrum, accurate prediction of the {\bf total} 
integrated semileptonic decay rate can be obtained \cite{hqet} within the HQET
including the first non-trivial non-perturbative
corrections as well as radiative perturbative QCD correction
\cite{kuhn}. The related uncertainties in calculation of the integrated
decay rate have been also analyzed \cite{luke,shifman,ball}.
The total inclusive semileptonic decay rate for $B \rightarrow X_q l \nu$ is
given  as
\begin{eqalignno}
\Gamma (B &\longrightarrow X_q l \nu) = 
{G_F^2 m_b^5 \over 192 \pi^3} |V_{qb}|^2 \times \label{eq2}\\
\Biggl\{\Biggl[z_0(x_q) &- {2 \alpha_s(m_b^2) \over 3 \pi} g(x_q) \Biggr] 
\left( 1 - {\mu_\pi^2 - \mu_G^2 \over 2 m_b^2} \right) \nonumber\\
&- z_1(x_q) {\mu_G^2 \over m_b^2} + 
{\cal O}(\alpha_s^2,\alpha_s/m_b^2,1/m_b^3)
~~\Biggr\} \nonumber
\end{eqalignno}
where
\begin{eqalignno}
x_q &\equiv m_q/m_b~~, \nonumber\\
z_0(x) &= 1 -8x^2 +8x^6 -x^8 -24x^4\log{x}~~, \nonumber\\
z_1(x) &= (1-x^2)^4~~, \nonumber
\end{eqalignno}
and $g(x) = (\pi^2-31/4)(1-x)^2+3/2$
is the corresponding single gluon
exchange perturbative QCD correction \cite{kuhn,kim-martin}.

The expectation value of energy due to the chromomagnetic hyperfine 
interaction, $\mu_G$, can be related to the $B^* - B$ mass difference
\begin{eqalignno}
\mu_G^2 = {3 \over 4} (M_{B^*}^2 - M_B^2) 
\approx (0.35 \pm 0.005) {\rm GeV}^2
\label{eq4}
\end{eqalignno}
and the expectation value of kinetic energy of $b$-quark inside the $B$ meson, 
$\mu_\pi^2$, is given from various  
arguments \cite{mu-pi,kim-namgung,gremm},
\begin{eqalignno}
0.1~{\rm GeV}^2 \leq \mu_\pi^2 \leq  0.7~{\rm GeV}^2,
\label{eq5}
\end{eqalignno}
which shows much larger uncertainties compared to $\mu_G^2$.
The value of $|V_{cb}|$ has been estimated \cite{luke,shifman,ball}
from the total decay rate $\Gamma(B \rightarrow X_c l \nu)$ 
of Eq. (\ref{eq2}) by using the pole mass of 
$m_b$ and a mass difference $(m_b - m_c)$ based on the HQET. 
As can be easily seen from Eq. (\ref{eq2}), the $m_b^5$ factor,
which appears in the semileptonic decay rate, 
but not in the branching fraction, 
is the largest source of the uncertainty, resulting in about
$5 \sim 20\%$ error in the prediction of $|V_{cb}|$ via the semileptonic 
branching fraction and $B$ meson life time \cite{luke,shifman,ball}.
Historically, the ACCMM model \cite{accmm} was motivated to avoid this  
$m_b^5$ factor, and at the same time to naively incorporate the bound state 
effect of initial $B$ meson.

We can do a similar exercise to predict the value of $|V_{ub}|$ from the
integrated total decay rate of $\Gamma(B \rightarrow X_u l \nu)$, to find out
\begin{eqalignno}
|V_{ub}|^2 = {192 \pi^3 \over G_F^2 m_b^5} &\cdot 
\Gamma(B \rightarrow X_u l \nu) \\ 
\times \Biggl\{ \Biggl[ 1 &- {2 \alpha_s(m_b^2) \over 3 \pi}
\left( \pi^2 - {25 \over 4} \right) \Biggr]  \nonumber\\
&\cdot \Biggl( 1 - {\mu_\pi^2 - \mu_G^2 \over 2 m_b^2} \Biggr) 
- {\mu_G^2 \over m_b^2} \Biggr\}^{-1}. \nonumber
\end{eqalignno}
We use the pole mass of $b$-quark $m_b = (4.8 \pm 0.2)$ GeV 
from a QCD sum-rule analysis 
of the $\Upsilon$-system \cite{voloshin}.
To be conservative, 
we use here a larger error bar (larger by a factor 8) than that 
of the original analysis \cite{voloshin}. 
We estimate the largest possible error of $m_b$ as ${\cal O}(\Lambda_{QCD})$. 
And  $x_u \equiv m_u/m_b \simeq 0$
and we take $\alpha_s(m_b^2) = (0.24 \pm 0.02)$. 
[Extrapolating the known 5 \% error of 
$\alpha_s(m_{_Z}^2)$, we estimate about 10 \% error for $\alpha_s(m_b^2)$.]

We get numerically
\begin{eqalignno}
\gamma_{u} &\equiv
{\Gamma_{theory}(B \rightarrow X_u l \nu) \over |V_{ub}|^2} \nonumber\\
&\simeq (7.1 \pm 1.5) \times 10^{13}/sec, \nonumber
\end{eqalignno}
and
\begin{eqalignno}
|V_{ub}| &\simeq (3.6 \pm 0.4) \times 10^{-3}  \label{eq6} \\ &\times
\left[{{\cal B}(B \rightarrow X_u l \nu) \over 1.4\times 10^{-3}}\right]^{1/2}
\left[ { 1.52~ psec \over \tau_{_B} } \right]^{1/2}. \nonumber
\end{eqalignno}

As previously explained, the largest uncertainty comes from the factor $m_b^5$,
which gives the most part of the theoretical errors shown in Eq. (\ref{eq6}).
We remark that the semileptonic branching fraction of $b \rightarrow u$ decay,
${\cal B}(B \rightarrow X_u l \nu)$, has to be precisely measured 
to experimentally determine the value of $|V_{ub}|$ from Eq. (\ref{eq6}). 
We will discuss on the
experimental possibilities in details in the next Section.
Once the inclusive branching fraction ${\cal B}(B \rightarrow X_u l \nu)$ is 
precisely measured, we can extract the value of $|V_{ub}|$ within the 
theoretical error ($\sim 10 \%$) similar to those of $|V_{cb}|$. 
(Compare the inclusive $\gamma_{u}$ of Eq. (\ref{eq6}) and the exclusive 
$\gamma_{\rho}$ of Eq. (\ref{eq1}) for the theoretical predictions of 
the semileptonic $b \rightarrow u$ decay.)

The ratio of CKM matrix elements $|V_{ub}/V_{cb}|$  can be determined in 
a model-independent way by taking the ratio of semileptonic decay widths
$\Gamma(B \rightarrow X_u l \nu)/\Gamma(B \rightarrow X_c l \nu)$.
As can be seen from Eq. (\ref{eq2}), 
this ratio is theoretically described by the 
phase space factor and the well-known perturbative QCD correction only,
\begin{eqalignno}
{\Gamma(B \rightarrow X_u l \nu) \over \Gamma(B \rightarrow X_c l \nu)}
&\simeq \left| { V_{ub} \over V_{cb} } \right|^2 
\Biggl[ 1 - {2 \alpha_s \over 3 \pi}
\left( \pi^2 - {25 \over 4} \right) \Biggr] \nonumber\\
&\times \Biggl[ z_0(x_c) - {2 \alpha_s \over 3 \pi} g(x_c) \Biggr]^{-1},
\label{eq7}
\end{eqalignno}
where we ignored the term $\mu_G^2/m_b^2$, which gives about
1 \% correction to the ratio.
We strongly emphasize here 
that the sources of the main theoretical uncertainties, 
the most unruly factor $m_b^5$ and the still-problematic non-perturbative 
contributions, are all
canceled out in this ratio. By taking $\alpha_s(m_b^2) = (0.24 \pm 0.02)$,
and by using the mass difference relation from the HQET \cite{mass}, 
which gives $x_c \equiv m_c/m_b \approx 0.25 - 0.30$.
[This ratio $x_c$ is calculable from the mass difference
$(m_b-m_c)$, which also includes the uncertain parameter 
$\mu_\pi^2$ of Eq. (\ref{eq5}) as a small correction factor.]
 
The ratio of the semileptonic decay widths is estimated as
\begin{eqalignno}
{\Gamma(B \rightarrow X_u l \nu) \over \Gamma(B \rightarrow X_c l \nu)}
&\equiv \left({\gamma_u \over \gamma_c}\right)\times 
   \left| { V_{ub} \over V_{cb} } \right|^2 \label{eqq8}\\
&\simeq (1.83 \pm 0.28) 
	\times \left| { V_{ub} \over V_{cb} } \right|^2, \nonumber
\end{eqalignno}
and the ratio of CKM elements is 
\begin{eqalignno}
\left| { V_{ub} \over V_{cb} } \right| 
&\equiv \left({\gamma_c \over \gamma_u}\right)^{1/2} \times
\left[ {{\cal B}(B \rightarrow X_u l \nu) \over 
{\cal B}(B \rightarrow X_c l \nu) } \right]^{1/2} \label{eq8}\\
&\simeq (0.74 \pm 0.06) \times
\left[ {{\cal B}(B \rightarrow X_u l \nu) \over 
{\cal B}(B \rightarrow X_c l \nu) } \right]^{1/2}. \nonumber
\end{eqalignno}

Once the ratio of semileptonic decay widths (or equivalently the ratio of
branching fractions 
${\cal B}(B \rightarrow X_u l \nu)/{\cal B}(B \rightarrow X_c l \nu)$)
is measured in the forthcoming asymmetric $B$-factories, 
this should give a powerful model-independent determination 
of $|V_{ub}/V_{cb}|$.
There is absolutely no theoretical model dependence in these ratios,
Eqs. (\ref{eq7},\ref{eqq8},\ref{eq8}).
As explained earlier, for example, in the ACCMM model \cite{accmm} the model 
dependence comes in via the introduction of the parameter $p_{_F}$ 
in place of the factor $m_b^5$, which is now canceled in these ratios.
We also note that by using the integrated total decay widths, instead of 
the lepton energy spectrum, another possible model dependence related to the 
end point region of the spectrum need not be even introduced. 

\subsection{Experimental possibility}

As explained in the previous Section,
in order to measure $|V_{ub}/V_{cb}|$ (and $|V_{ub}|$) model-independently
by using the relations Eqs. (\ref{eq6},\ref{eq8}), it is experimentally
required to separate the $b \rightarrow u$ semileptonic decays 
from the dominant $b \rightarrow c$ semileptonic decays, 
and to precisely measure the branching fraction
${\cal B}(B \rightarrow X_u l \nu)$ or the ratio
${\cal B}(B \rightarrow X_u l \nu)/{\cal B}(B \rightarrow X_c l \nu)$.
At presently existing symmetric $B$-experiments, ARGUS and CLEO, where $B$ and 
$\bar B$ are produced almost at rest, this required separation is possible only
in the very end-point region of the lepton energy spectrum, because both $B$ 
and $\bar B$ decay into the whole $4 \pi$ solid angle from the almost same 
decay point, and it is not possible to identify the parent $B$ meson of 
each produced particle. Hence all the hadronic information is of no use.
However, in the forthcoming asymmetric $B$-experiments with microvertex 
detectors, BABAR and BELLE \cite{B-fact}, where the two
beams have different energies and the produced $\Upsilon(4S)$ is not at
rest in the laboratory frame, the bottom decay vertices will be
identifiable.
The efficiency for the full reconstruction of each event could be 
relatively high limited 
only by the $\pi^0$-reconstruction efficiency of about $60 \%$ \cite{B-fact}, 
and this $b \rightarrow u$ separation would be experimentally viable.

As of the most straightforward separation method, the measurements of 
inclusive hadronic invariant mass ($m_{_X}$) distributions in 
$B \rightarrow X_{c,u} l \nu$   can be very useful 
for the fully reconstructed semileptonic decay events.  
For $b \rightarrow c$ decays, one necessarily has  
$m_{_X} \geq m_{_D} = 1.86$ GeV.  
Therefore, if we impose a condition $m_{_X} < m_{_D}$, 
the resulting events come only from $b \rightarrow u$ decays,  
and about 90\% of the $b\rightarrow u$ events would survive this cut.   
This is already in sharp contrast with the usual cut 
on charged lepton energy $E_l$.

In fact, one can relax the condition  $m_{_X} < m_{_D}$, and extract almost
the total $b \rightarrow u$ semileptonic decay rate \cite{kim-ko,cskim}, 
because the $m_{_X}$ distribution in $b \rightarrow c$ decays is completely 
dominated by contributions of three resonances $D, D^{*} $ and $D^{**}$, 
which are  essentially like $\delta$-functions, 
\begin{eqalignno} 
{d \Gamma \over d m_{_X}} = \Gamma(B\rightarrow R l \nu)~
\delta(m_{_X} - m_{_R})~~,
\label{eq9}
\end{eqalignno}
where the resonance $R = D, D^*$ or $D^{**}$. See Fig. 1.
In other words, one is allowed to use the $b \rightarrow u$ events 
in the region even above $m_{_X} \geq m_{_D}$, first by excluding small 
regions in $m_{_X}$ around $m_{_X} = m_{_D}, m_{_{D^{*}}}, m_{_{D^{**}}}$,
and then by including the regions again numerically 
in the $m_{_X}$ distribution of $b \rightarrow u$ decay from its values just
around the resonances. 
There still is a non-resonant decay background at large invariant-mass region
$m_{_X} \geq m_{_D} + m_\pi$ from $B \rightarrow (D + \pi) l \nu$ in using 
this inclusive $m_{_X}$ distribution separation.
However, with an additional `$D - \pi$' separation condition, which we
explain later, this many-pion producing  $b \rightarrow u$ decay
could be safely differentiated from the non-resonant $b \rightarrow c$ 
semileptonic decays.
Instead of using the `$D - \pi$' separation,
if we impose a condition $m_{_X} < m_{_D} + m_\pi$, we would get
about 95\% of the total $b\rightarrow u$ events.

We note that there is possibly a question of bias.  Some classes of final
states ($e.g.$ those with low multiplicity, few neutrals) may be more
susceptible to a full and unambiguous reconstruction. Hence an
analysis that requires this reconstruction may be biased. However,
the use of topological information from microvertex detectors
should tend to reduce the bias, since vertex resolvability depends
largely on the proper time of the decay and its orientation relative
to the initial momentum (that are independent of the decay mode).
Also such a bias can be allowed for in the analyses, via a suitable
Monte Carlo modeling.
For more details on this inclusive hadronic invariant mass distribution
$d \Gamma / d m_{_X}$, please see Ref. \cite{kim-ko}.

Even without full reconstructions of all the final particles, 
one can separate $b \rightarrow u$ decays from $b \rightarrow c$ decays
by using the particle decay properties \cite{pdg}.  Since 
$D^{**} \rightarrow D^* + \pi$ and $D^* \rightarrow D + \pi$, the semileptonic
$b \rightarrow c$ decays always produce at least one final state $D$ meson,
compared to $b \rightarrow u$ decays which produce particles, $\pi,~\rho,~...$
that always decay to one or more $\pi$ mesons at the end. 
Therefore, the $b \rightarrow u$ decay separation could be achieved better 
with the accurate `$D - \pi$' separation in particle detectors, 
and if combined with the hadronic invariant-mass distribution separation. 
Such `$D - \pi$' separations would be possible via partial reconstructions
of a whole event or by using special event characteristics,
rather than by fully reconstructing a whole event.
In order for this 
`$D - \pi$' separation to be successfully implemented,
one should have the separation with better than about 98 \% efficiency, 
because of $\Gamma(B \rightarrow X_u l \nu) \sim (0.02) \times 
\Gamma(B \rightarrow X_c l \nu)$.
We can also think of `$K - \pi$' separation as a natural extension,
because the semileptonic
$b \rightarrow c$ decays always produce at least one final state $K$ meson.
These `$D - \pi$' (and `$K - \pi$') separation conditions are also
applicable at the hadronic $B$-factories, HERA-B, LHC-B.

There  possibly is a source of background to this `$D - \pi$' 
separation condition from the cascade decay of 
$b \rightarrow c \rightarrow s l \nu$. Recently ARGUS and CLEO \cite{cas-B} 
have separated this cascade decay background
from the signal events to extract the model-independent spectrum of 
${d\Gamma \over dE_l}(B \rightarrow X_c l \nu)$ for the whole region of 
electron energy, by taking care of lepton charge and $B - \bar B$ mixing 
systematically. In the future asymmetric $B$-factories with much higher 
statistics, this cascade decay may not be any serious background at all
except for the case with very low energy electron production. 

\section{DISCUSSIONS \& CONCLUSIONS}

The precise value of $V_{ub}$ is urgently needed 
for understanding the origin of CP-violation, for
testing the SM through the unitarity relations
among them, and also in the searches for hints of new physics beyond the SM.
We propose that the ratio of CKM matrix elements $|V_{ub}/V_{cb}|$  
can be determined in 
a model-independent way by taking the ratio of semileptonic decay widths
$\Gamma(B \rightarrow X_u l \nu)/\Gamma(B \rightarrow X_c l \nu)$, which
is theoretically described by the phase space factor
and the well-known perturbative QCD correction only,
and which predicts
\begin{eqalignno}
{\Gamma(B \rightarrow X_u l \nu) \over \Gamma(B \rightarrow X_c l \nu)}
&\equiv \left({\gamma_u \over \gamma_c}\right) \times 
\left| {V_{ub} \over V_{cb}}\right|^2 \nonumber\\
&\simeq (1.83 \pm 0.28) \times \left|{V_{ub} \over V_{cb}}\right|^2, \nonumber
\end{eqalignno}
and
\begin{eqalignno}
\left|{V_{ub} \over V_{cb}}\right| &\equiv 
\left({\gamma_c \over \gamma_u}\right)^{1/2} \times 
\left[{{\cal{B}}(B \rightarrow X_u l \nu) \over
{\cal{B}}(B \rightarrow X_c l \nu)}\right]^{1/2} \nonumber\\
&\simeq (0.74 \pm 0.06) \times \left[{{\cal{B}}(B \rightarrow X_u l \nu) \over
{\cal{B}}(B \rightarrow X_c l \nu)}\right]^{1/2}, \nonumber
\end{eqalignno}
based on the heavy quark effective theory. 
Once the ratio of semileptonic decay widths 
(or equivalently the ratio of branching fractions 
${\cal B}(B \rightarrow X_u l \nu)/{\cal B}(B \rightarrow X_c l \nu)$)
is measured, this ratio will give a powerful 
model-independent determination of $|V_{ub}/V_{cb}|$.

In the forthcoming asymmetric $B$-factories with microvertex 
detectors, the total separation of $b \rightarrow u$ 
semileptonic decays from the dominant $b \rightarrow c$ semileptonic decays 
to determine the ratio would be experimentally viable.
We explore the possible experimental options: the measurement of
inclusive hadronic invariant mass distributions,  and
the `$D - \pi$' (and `$K - \pi$') separation conditions.
We also clarify the relevant experimental backgrounds.
In view of the potential importance of 
${\cal B}(B \rightarrow X_u l \nu )/{\cal B}(B \rightarrow X_c l \nu )$ 
as a new theoretically model-independent probe for measuring
$|V_{ub}/V_{cb}|$,
we would like to urge our experimental colleagues to make sure that 
this $b \rightarrow u$ separation can indeed be successfully achieved.

\begin{figure}[htb]
\vspace{9pt}
\hbox {\epsfig{file=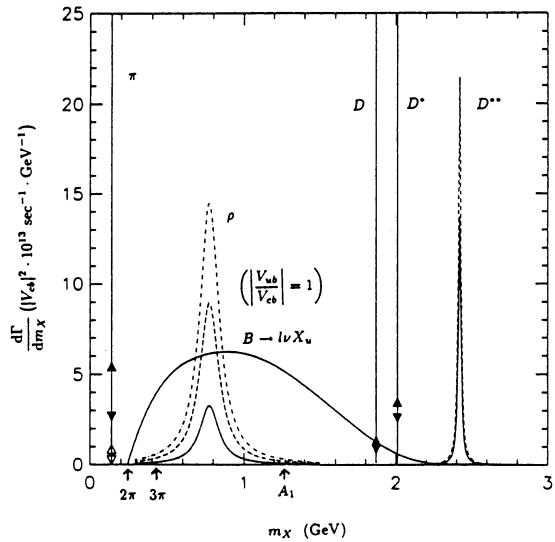}}
\caption{The $m_X$ distributions  in $B \rightarrow X_{c,u} l \nu$
with $| V_{ub} / V_{cb} | = 1$.  The $b \rightarrow c$ transition is
dominated by the $X_{c} = D, D^{*}, D^{**}$.  
On the other hand,  the $b \rightarrow u$ transition is largely nonresonant.
The cases with $X_{u} = \pi, \rho$ are shown explicitly. 
The inclusive $m_{X}$ distribution for
$b \rightarrow u$ was obtained from the ACCMM model  
with hadronic mass constraint of $m_X \stackrel{>}{\sim} 2 m_{\pi}$.
Note that the individual exclusive decay width, 
$e.g.$ $\gamma_\rho$ as in Eq (1),
depends strongly on the models, even though the total inclusive decay rate
is calculable model-independently within the HQET.}
\label{fig:largenenough}
\end{figure}

\end{document}